\documentclass[12pt]{article}
\usepackage{cite}
\usepackage{enumerate}
\usepackage{ifpdf}
\usepackage{ae}
\usepackage[T1]{fontenc}
\usepackage[ansinew]{inputenc}
\usepackage{amsmath}
\usepackage{relsize}
\usepackage{amssymb}
\usepackage{tikz}
\usepackage{graphicx}
\usepackage{color}
\definecolor{darkblue}{cmyk}{0.9,0.9,0,0}
\definecolor{darkgreen}{rgb}{0,0.55,0}
\usepackage[colorlinks=true,linkcolor=darkblue,citecolor=darkblue,urlcolor=darkblue]{hyperref}
\usepackage{epsfig}
\usepackage{epstopdf}
\usepackage{graphicx}

\makeatletter
\long\def\@makecaption#1#2{
  \vskip\abovecaptionskip
  \sbox\@tempboxa{{\captionfonts #1: #2}}
  \ifdim \wd\@tempboxa >\hsize
    {\captionfonts #1: #2\par}
  \else
    \hbox to\hsize{\hfil\box\@tempboxa\hfil}
  \fi
  \vskip\belowcaptionskip}
\makeatother

\newcommand{\beq}{\begin{equation}}
\newcommand{\eeq}{\end{equation}}
\newcommand{\beqy} {\begin{eqnarray}}
\newcommand{\eeqy} {\end{eqnarray}}
\newcommand{\bsmat}{\begin{smallmatrix}}
\newcommand{\esmat}{\end{smallmatrix}}
\newcommand{\bmat}{\begin{matrix}}
\newcommand{\emat}{\end{matrix}}

\def\({\left(}
\def\){\right)}
\def\[{\left[}
\def\]{\right]}

\def\<{\langle}
\def\>{\rangle}

\usepackage{varioref}
\usepackage{makeidx}
\makeindex

\usepackage[english]{babel}
\usepackage{caption}
\usepackage{subfig}

\usepackage{tabularx}
\begin{document}

\thispagestyle{empty}

\renewcommand{\thefootnote}{\fnsymbol{footnote}}
\setcounter{page}{1}
\setcounter{footnote}{0}
\setcounter{figure}{0}

\begin{titlepage}

\begin{center}

\vskip 2.3 cm 

\vskip 5mm

{\Large \bf
Unitarity and positivity constraints for CFT at large central charge
}
\vskip 0.5cm

\vskip 15mm

\centerline{Luis F. Alday$^a$ and Agnese Bissi$^b$  }
\bigskip
\centerline{\it $^a$ Mathematical Institute, University of Oxford,} 
\centerline{\it Woodstock Road, Oxford, OX2 6GG, UK}
\centerline{\it $^b$ Center for the Fundamental Laws of Nature,}
\centerline{\it Harvard University, Cambridge, MA 02138 USA}

\end{center}

\vskip 2 cm

\begin{abstract}
\noindent We consider the four-point correlator of the stress tensor multiplet in ${\cal N}=4$ SYM in the limit of large central charge $c \sim N^2$. For finite values of $g^2N$ single-trace intermediate operators arise at order $1/c$ and this leads to specific poles in the Mellin representation of the correlator. The sign of the residue at these poles is fixed by unitarity. We consider solutions consistent with crossing symmetry and this pole structure. We show that in a certain regime all solutions result in a negative contribution to the anomalous dimension of twist four operators. The reason behind this is a positivity property of Mack polynomials that leads to a positivity condition for the Mellin amplitude. This positivity condition can also be proven by assuming the correct Regge behaviour for the Mellin amplitude. For large $g^2N$ we recover a tower of solutions in one to one correspondence with local interactions in a effective field theory in the $AdS$ bulk, with the appropriate suppression factors, and with definite overall signs. These signs agree with the signs that would follow from causality constraints on the effective field theory. The positivity constraints arising from CFT for the Mellin amplitude take a very similar form to the causality constraint for the forward limit of the S-matrix.\end{abstract}

\end{titlepage}


\setcounter{page}{1}
\renewcommand{\thefootnote}{\arabic{footnote}}
\setcounter{footnote}{0}

 \def\nref#1{{(\ref{#1})}}

\section{Introduction}

Invigorated by \cite{Rattazzi:2008pe}, the methods from the conformal bootstrap have led, over recent years, to significative progress in the understanding of conformal field theories in dimensions higher than two. The conformal bootstrap program consists in the study of conformal field theory from consistency conditions, namely, the structure of the operator algebra and crossing symmetry, unitarity and symmetries of the theory. This has led to a wide spectrum of results for the CFT data, both numeric as well as analytic. 

An interesting question is which conformal field theories admit a holographic dual, and for those who do, how does the geometry emerge from the CFT. Conformal bootstrap methods where used in \cite{Heemskerk:2009pn} in order to argue that any conformal field theory with a large $N$ expansion and a parametrically large gap in the spectrum of anomalous dimensions admits a dual local bulk theory. This was further developed in \cite{Heemskerk:2010ty,Fitzpatrick:2010zm, Penedones:2010ue, Fitzpatrick:2011ia, Fitzpatrick:2011dm, Fitzpatrick:2011hu, Fitzpatrick:2012cg, Alday:2014tsa} . In the simplest set-up it is assumed that the spectrum at $N=\infty$ contains a single-trace scalar operator of dimension $\Delta$, and double trace operators of dimension $\Delta_{n,\ell} = 2\Delta+2n+\ell$, while any other single trace operators have a parametrically large dimension $\sim \Delta_{gap}$. The authors then consider the four point correlator of the single trace operator and construct solutions consistent with crossing symmetry and the structure of the OPE to order $1/N^2$. These solutions are then shown to be in one to one correspondence with local interactions in a scalar effective field theory in the bulk. 

A CFT that satisfies the conditions above is ${\cal N}=4$ SYM. In ${\cal N}=4$ SCFT the stress energy tensor sits in a half-BPS multiplet, whose superconformal primary is a single trace operator ${\cal O}_{\bf 20'}$ of dimension two. The crossing relation for the four-point correlator can be expressed as a relation for the intermediate {\it unprotected} operators present in the OPE ${\cal O}_{\bf 20'} \times {\cal O}_{\bf 20'}$, see \cite{Beem:2013qxa}. At $N=\infty$ the spectrum of unprotected operators contains towers of double trace operators as described above. Furthermore at large $\lambda=g^2 N$ all unprotected single trace operators acquire a large scaling dimension. While this theory has a well known gravity dual, it is an interesting question how much of the structure of this dual theory can be recovered purely from CFT considerations. In \cite{Alday:2014tsa} we applied the ideas of \cite{Heemskerk:2009pn}  to ${\cal N}=4$ SYM and constructed four-point correlators consistent with crossing symmetry and the structure of the OPE at order $1/N^2$. In this case, the solutions must contain the {\it supergravity} term. In addition, crossing allows for a tower of solutions again in one to one correspondence with interactions in a bulk effective field theory. 

Consider a generic bulk effective field theory of a massless scalar field in $AdS_{d+1}$

\begin{equation}
S = \int d^{d+1}x \sqrt{-g} \left( - \partial_\mu \varphi \partial^\mu \varphi  + \mu_1 \varphi^4 + \mu_2 \left( \partial_\mu \varphi \partial^\mu \varphi \right)^2 + \cdots\right)
\end{equation}
As already mentioned, each interaction term corresponds to a solution to the crossing conditions of the CFT at order $1/N^2$. Furthermore, each term will lead to a contribution, proportional to $\mu_1,\mu_2$, etc, to the anomalous dimensions of double trace operators:

\begin{equation}
\Delta_{n,\ell} = 2\Delta+ 2n + \ell + \frac{1}{N^2} \gamma_{n,\ell}
\end{equation}
 From the point of view of an $AdS$ effective field theory the coefficients $\mu_1,\mu_2$, etc are not unconstrained. On one hand, non-renormalizable interactions are expected to be suppressed by an extra scale, which should correspond to $\Delta_{gap}$, and the precise power of $\Delta_{gap}$ should follow from dimensional analysis, see \cite{Fitzpatrick:2010zm, Fitzpatrick:2012cg}. On the other hand causality constraints enforce a specific sign for some of the couplings. For instance, unless $\mu_2 \geq 0$ the effective theory cannot be UV-completed, see \cite{Adams:2006sv}. This constraint will then lead to a corresponding constraint for the {\it sign} of the contribution to the anomalous dimension  $\gamma_{n,\ell}$ proportional to $\mu_2$. 
 
From the point of view of the CFT, however, the coefficient in front of each solution to the crossing equations (and its sign), is unconstrained. In \cite{Alday:2014tsa} it was shown that the constraints arising from the point of view of effective field theory, together with the constraints from causality obtained in \cite{Adams:2006sv}, lead to either a negative or suppressed correction to the anomalous dimension of twist four operators, in the regime of large $\lambda$. This is consistent with the numeric bounds found in \cite{Beem:2013qxa}. Furthermore, it was also seen that the constraints arising from causality alone did not necessarily imply a negative contribution to the anomalous dimension of all twist four operators. 

At any extent, a purely CFT handle of the coefficients $\mu_1,\mu_2,$ etc, is hard to achieve. Part of the problem is that in the approaches of  \cite{Heemskerk:2009pn} and \cite{Alday:2014tsa}  the constraints of crossing symmetry are implemented, but the constraints of unitarity are lost at order $1/N^2$. \footnote{Certain progress can be made by resorting to the concept of perturbative unitarity, see \cite{Fitzpatrick:2012cg}.} In a beautiful paper, Hartman and collaborators  have analysed causality constraints in conformal field theories \cite{Hartman:2015lfa}. In particular, they have shown that $\mu_2 \geq 0$ also follows from CFT considerations. This in turn leads to the correct sign for the anomalous dimensions of twist four, spin two, double trace operators. 

In this paper we make the following simple observation. In the treatment \cite{Alday:2014tsa}, it was assumed that the dimension of all unprotected single trace operators was very large. In ${\cal N}=4$ SYM this will happen for large $\lambda$. In the  present paper we relax this assumption and assume that single trace operators arise in the OPE ${\cal O}_{\bf 20'} \times {\cal O}_{\bf 20'}$ at order $1/N^2$. Due to unitarity such new contributions appear with a definite positive sign. We find it convenient to work in Mellin space \cite{Penedones:2010ue, Fitzpatrick:2011ia, Paulos:2011ie}, with the correlator given by

\begin{equation}
{\cal A}(u,v) \sim  \int \Gamma^2(x+2) \Gamma^2(y+2) \Gamma^2(-x-y) M(x,y) u^{-x} v^{-y} dx dy
\end{equation}
The presence of a new intermediate operator of twist $\tau$ in the direct channel leads to extra single-poles in the Mellin expression for the correlator
\begin{equation}
M(x,y) \sim \frac{1}{x+\tau/2+n}
\end{equation}
We consider non-polynomial solutions consistent with crossing symmetry and the correct analytic structure of the Mellin representation. We then study the contribution from such solutions to the anomalous dimensions of double trace twist four operators. We show that for intermediate twist higher than four, $\tau>4$, the contribution is always negative. The basic reason behind this is certain positivity condition for Mack polynomials. This in turn leads to a positivity condition for a slice of the Mellin amplitude, in the regime in which the twist of all unprotected single trace operators is higher than four:
\begin{equation}
M(x,-2,-x) = \sum_{n=0,2,\cdots} c_n {\cal F}_n(x),~~~~~~c_n \geq 0.
\end{equation}
where ${\cal F}_n(x)$ are continuous Hahn polynomials. In constructing the non-polynomial solutions for higher spin exchange there is always the ambiguity of adding regular polynomial terms. These ambiguous terms may in principle spoil the positivity condition, however we show that the same positivity condition follows if we assume the Mellin amplitude has the correct Regge behaviour. This condition implies the anomalous dimension of double trace twist four operators is always negative and explains the numeric results observed in \cite{Beem:2013qxa}.

We then consider the limit of very large $\tau$. We show that in this limit the solutions corresponding to the exchange of single trace scalar operators give rise to the tower of polynomial solutions considered in  \cite{Alday:2014tsa}, with higher and higher order terms suppressed by powers of $\tau$. The structure of this expansion exactly agrees with the structure proposed in \cite{Alday:2014tsa} based on effective field theory. In addition, the overall signs are fixed, such that the contribution to the dimension of twist-four double trace operators is negative. Similar conclusions can be drawn for the exchange of operators with spin, assuming the polynomial ambiguities do not spoil the large twist limit. 
 
 It is instructive to consider the flat space limit of the Mellin amplitude. Causality constraints impose certain positivity conditions for the forward limit of the flat space S-matrix \cite{Adams:2006sv}. These constraints have a very similar form to the positivity constraints for the Mellin amplitude mentioned above. Finally, we consider the large $n$ behaviour of $\gamma_{n,\ell}$. The spectrum of operators should be such that in the flat space limit the Virasoro-Shapiro amplitude is recovered. This result can be used to write down the leading large $n$ behaviour of $\gamma_{n,\ell}$ to all orders in $1/\sqrt{\lambda}$. A generic growth in $\gamma_{n,\ell}$ leads to a divergence in the Lorentzian correlator. This divergence signals the locality of the bulk theory. By using the above behaviour for $\gamma_{n,\ell}$ a leading order expression for such divergence is given. 

This paper is organised as follows. In section two we consider a family of solutions consistent with crossing symmetry and the exchange of single trace operators. We show that these solutions lead to a negative contribution to the anomalous dimension of twist four operators in a specific regime. The reason behind this is certain positivity property for Mack polynomials that lead to a positivity condition for a slice of the Mellin amplitude. We then show that the same condition follows from requiring the correct Regge behaviour. In section three we consider several instructive limits of the Mellin amplitude and the solutions of section two. Namely, the large twist limit of the non-polynomial solutions and the flat space limit of the Mellin amplitude. Moreover, we study the large $n$ behaviour of the anomalous dimensions of double trace operators and the Lorentzian singularities this leads to. We end with some conclusions and open problems. We defer some technical details to the appendices. 

\section{Analytic solutions at large $N$ and single trace operators}

\subsection{Generalities}
In this paper we will consider a specific correlator in four-dimensional ${\cal N}=4$ SYM. In this theory the stress tensor sits in a half-BPS multiplet, whose superconformal primary ${\cal O}_{\bf 20'}$ is a scalar operator of protected dimension two, which transforms in the ${\bf 20'}$ of the $SU(4)$ $R-$symmetry group. Conformal invariance implies

\begin{equation}
\langle {\cal O}_{\bf 20'}(x_1){\cal O}_{\bf 20'}(x_2){\cal O}_{\bf 20'}(x_3){\cal O}_{\bf 20'}(x_4) \rangle = \sum_{\cal R} \frac{{\cal G}^{({\cal R})}(u,v)}{x_{12}^4 x_{34}^4}
\end{equation}
where the sum runs over the six representations present in the tensor product ${\bf 20'} \times {\bf 20'}$ and we have introduced the standard cross ratios
\begin{equation}
u= \frac{x_{12}^2 x_{34}^2}{x_{13}^2 x_{24}^2} = z \bar z,~~~~~~v= \frac{x_{14}^2 x_{23}^2}{x_{13}^2 x_{24}^2} = (1-z)(1- \bar z).
\end{equation}
Superconformal Ward identities relate the contributions from different representations ${\cal G}^{({\cal R})}(u,v)$ and allow us to write all contributions in terms of a single function ${\cal G}(u,v)$. The OPE ${\cal O}_{\bf 20'} \times {\cal O}_{\bf 20'}$ contains both, operators in long multiplets as well as operators in (semi-)short multiplets \cite{Nirschl:2004pa, Dolan:2004iy}. Consequently ${\cal G}(u,v)$ admits the decomposition

\begin{equation}
{\cal G}(u,v) = {\cal G}^{short}(u,v)+{\cal G}^{long}(u,v)
\end{equation}
where
\begin{equation}
\label{GlonCPW}
{\cal G}^{long}(u,v) = \sum_{\Delta,\ell} a_{\Delta,\ell} u^{\frac{\Delta-\ell}{2}} g_{\Delta+4,\ell}(u,v).
\end{equation}
The sum runs over unprotected superconformal primary operators, singlet of $SU(4)$, with even Lorentz spin $\ell$ and scaling dimension $\Delta$. $a_{\Delta,\ell}$ denotes the square of the OPE coefficients. The contribution from superconformal descendants is taken into account by the superconformal blocks $u^{\frac{\Delta-\ell}{2}} g_{\Delta+4,\ell}(u,v)$, with
\begin{equation}
g_{\Delta,\ell}(u,v) = \frac{2^{-\ell}}{z - \bar z}\left( z^{\ell+1} k_{\Delta+\ell}(z) k_{\Delta-\ell-2}(\bar z) -\bar z^{\ell+1} k_{\Delta+\ell}(\bar z) k_{\Delta-\ell-2}(z)  \right)
\end{equation}
where we have introduced $k_\beta=~_2F_1(\beta/2,\beta/2,\beta;z)$. It is furthermore convenient to decompose ${\cal G}^{long}(u,v)$ into its Born approximation (free theory), which we denote by ${\cal G}^{long}_{Born}(u,v,N)$, plus a quantum contribution. The Born expression does not depend on the coupling, is explicitly known and it admits an expansion around $N=\infty$:

\begin{equation}
 {\cal G}^{long}_{Born}(u,v,N) =  {\cal G}^{long,(0)}_{Born}(u,v) + \frac{1}{N^2} {\cal G}^{long,(1)}_{Born}(u,v) +\cdots
\end{equation}
while the quantum correction vanishes in the limit $N=\infty$ and starts at order $1/N^2$. Hence we write
\begin{equation}
{\cal G}^{long}(u,v)  = {\cal G}^{long,(0)}_{Born}(u,v) + \frac{1}{N^2} {\cal G}^{long,(1)}_{Born}(u,v)+ \frac{1}{N^2} \frac{{\cal A}(u,v)}{v^2}  + \cdots
\end{equation}
Invariance of the full correlator under the exchange of any two operators leads to crossing symmetry relations. ${\cal G}^{long}_{Born}(u,v,N)$ mixes with ${\cal G}^{short}(u,v)$ while ${\cal A}(u,v)$ satisfies crossing relations by itself:
\begin{equation}
{\cal A}(u,v) = {\cal A}(v,u),~~~~~{\cal A}(u,v) = v^2 {\cal A}\left(\frac{u}{v},\frac{1}{v} \right)
\end{equation}
 At $N=\infty$ the space of intermediate states is spanned by double trace operators of the schematic form ${\cal O}_{n,\ell}=  {\cal O} \Box^n \partial_{\mu_1} \cdots  \partial_{\mu_\ell}  {\cal O}$, of spin $\ell$ and dimension $\Delta_{n,\ell}=4+2n+\ell$. The leading term ${\cal G}^{long,(0)}_{Born}(u,v)$ fixes the OPE coefficients at leading order:
 
 \begin{equation}
 a_{n,\ell}^{(0)}= \frac{\pi (1+\ell)(6+\ell+2n)\Gamma(3+n)\Gamma(4+\ell+n)}{2^{7+\ell+4n}\Gamma\left(\frac{5}{2}+n \right)\Gamma\left(\frac{7}{2}+\ell+n \right)}.
\end{equation}
Next we would like to consider the correlator in a large $N$ expansion and look for solutions consistent with crossing symmetry and the OPE expansion to order $1/N^2$. The dimensions and OPE coefficients of double trace operators admit an expansion
\begin{eqnarray}
\Delta_{n,\ell}&=& 4+2n+\ell + \frac{1}{N^2}\gamma_{n,\ell}+ \cdots , \\
a_{n,\ell} &=& a^{(0)}_{n,\ell} +\frac{1}{N^2}a^{(1)}_{n,\ell}  + \cdots .
\end{eqnarray}
The $1/N^2$ term ${\cal G}^{long,(1)}_{Born}(u,v)$ has a contribution which for small $u$ behaves as
 \begin{equation}
 {\cal G}^{long,(1)}_{Born}(u,v)= 16 u \frac{1-v^2+2v \log v}{v (1-v)^2} + \cdots
\end{equation}
however, for $\lambda \neq 0$ the CPW decomposition (\ref{GlonCPW}) does not contain operators of twist two. Hence, there should be a corresponding term in ${\cal A}(u,v)$ which cancels this contribution. This is given by the supergravity result ${\cal A}_{sugra}(u,v)=-16 u^2 v^2 \bar D_{2422}(u,v)$, where the $\bar D$-functions are defined for instance in \cite{D'Hoker:1999pj}.  This leads to the following anomalous dimension and correction to the OPE coefficients of the double trace operators 

\begin{eqnarray}
& \gamma^{sugra}_{n,\ell} = -\frac{4(1+n)(2+n)(3+n)(4+n)}{(1+\ell)(6+\ell+2n)}\\
& a^{(1),sugra}_{n,\ell} = \frac{1}{2} \frac{\partial}{\partial n} \left(a^{(0)}_{n,\ell}  \gamma^{sugra}_{n,\ell} \right)
\end{eqnarray}
In addition we can add any solution to the homogeneous crossing relations consistent with the structure of the CPW decomposition. In order to proceed, we will assume that the correlator at order $1/N^2$ admits a Mellin representation  \cite{Penedones:2010ue, Fitzpatrick:2011ia, Paulos:2011ie}

\begin{equation}
{\cal A}(u,v) = \frac{1}{(2\pi i)^2} \int \Gamma^2(x+2) \Gamma^2(y+2) \Gamma^2(-x-y) M(x,y) u^{-x} v^{-y} dx dy
\end{equation}
where the integration contours are over the imaginary axis shifted by a small positive real part. The crossing relations in Mellin space simply read

\begin{equation}
M(x,y) = M(y,x) = M(x,-2-x-y)
\end{equation}
It is convenient to introduce an extra variable $z$, such that $x+y+z=-2$. Crossing symmetry then implies that the Mellin expression is completely symmetric under permutation in the variables $(x,y,z)$.  In this language the supergravity solution corresponds to
\begin{equation}
M_{sugra}(x,y,z) = - \frac{16}{(x+1)(y+1)(z+1)}
\end{equation}
The extra tower of solutions considered in \cite{Alday:2014tsa} is simply given by completely symmetric polynomials in the variables $x,y,z$. Note that the prefactor in the definition of the Mellin expression contains poles at $x=-2,-3,\cdots$, which correspond to the twist of double trace operators.  A polynomial solution $M(x,y)$ will not add new poles but it will change the residues of the poles corresponding to double twist operators, giving a contribution to their anomalous dimension and OPE coefficients. 

\subsection{Single trace operators}
When considering $1/N^2$ corrections to a four-point correlator, two things can happen. 1.- The dimension and the OPE coefficients of the double trace operators acquire a correction of order $1/N^2$; 2.- New, single trace, operators may appear in the OPE. The new operators will enter with their classical dimension at leading order, since their OPE coefficient is already of order $1/N^2$. In ${\cal N}=4$ SYM at large $N$ and finite $\lambda$, both things happen. 

Let us assume that at order $1/N^2$ a new operator, of twist $\tau$ and spin $\ell$, arises in the OPE of the two external operators, with corresponding OPE coefficient $\frac{1}{N^2}  a_{\tau,\ell}$. An important point is that due to unitarity

\begin{equation}
a_{\tau,\ell} >0 .
\end{equation}
${\cal A}(u,v)$ should contain a term corresponding to the exchange of the new operator:
\begin{equation}
{\cal A}(u,v) = a_{\tau,\ell} v^2 u^{\tau/2} g_{\tau+4,\ell}(u,v) + \cdots
\end{equation}
While the expression in Mellin space should possess a corresponding pole

\begin{equation}
M_{\tau,\ell}(x,y) = \frac{h_{\tau,\ell}(y)}{(x+\tau/2)} + \cdots
\end{equation}
As shown in appendix \ref{singlepoleunitarity} having a single pole (or a finite number of them) for generic, not even integer $\tau$, is not consistent with unitarity. More precisely, one needs to include the tower of poles corresponding to the descendants of the new operator. At these poles

\begin{equation}
M_{\tau,\ell}(x,y) = \frac{h^{(k)}_{\tau,\ell}(y)}{(x+\tau/2+k)} + \cdots
\end{equation}
The residues at the poles are fixed by the proper expression for the super conformal blocks in Mellin space, given by

\begin{equation}
\label{cbmellin}
{\cal B}_{\Delta,\ell}(x,y,z) = i e^{-i \pi \Delta}(e^{i\pi(\ell-2x+\Delta)}-1) \frac{\Gamma(x-\frac{\ell+\Delta}{2}-2) \Gamma(x+\frac{\Delta-\ell}{2}) }{\Gamma^2(x+2)} P_{\Delta}^{(\ell)}(y,z)
\end{equation}
Where $ P_{\Delta}^{(\ell)}(y,z)$ is a symmetric polynomial of degree $\ell$, defined in appendix \ref{Mack}, and essentially coincides with the Mack polynomial.  Note that the second gamma function in the numerator has poles at the locations

\begin{equation}
x = -\frac{\Delta-\ell}{2}-k
\end{equation}
Corresponding to the primary operator plus all its descendants. Given (\ref{cbmellin}) we find:

\begin{equation}
\label{alpha}
h^{(k)}_{\tau,\ell}(y) = a_{\tau,\ell} \frac{2 (-1)^{k+1} \sin (\pi  \tau ) \Gamma (-\ell-k-\tau -2) }{\Gamma (k+1) \Gamma \left(-k-\frac{\tau }{2}+2\right)^2} \left.P_{\ell+\tau}^{(\ell)}(y,z)\right|_{x=-\frac{\Delta-\ell}{2}-k}
\end{equation}

\subsubsection{Single scalar primary}
Let us start with the simplest example of a scalar primary operator of dimension $\delta$. The corresponding Mellin representation contains a pole:

\begin{equation}
M_\delta(x,y) = \frac{h^{(0)}_{\delta,0}}{x+\delta/2} + \cdots
\end{equation}
The minimal solution consistent with crossing symmetry and the pole structure is
\begin{equation}
M^{(min)}_\delta(x,y) = h^{(0)}_{\delta,0} \left(\frac{1}{x+\delta/2}+\frac{1}{y+\delta/2} +\frac{1}{z+\delta/2}\right)
\end{equation}
where recall $x+y+z=-2$. We are interested in computing the contribution from such a solution to the anomalous dimension of twist four operators. Here we will follow a brute force approach: we will compute the corresponding solution in space-time and then perform a CPW expansion. Let us call the space time expression $A^{(min)}_\delta(u,v)$. We are interested in the terms proportional to $u^2 \log u$ in a small $u$ expansion:

\begin{equation}
A^{(min)}_\delta(u,v) = u^2 \log u~h_2(v) + \cdots
\end{equation}
Plugging $M_\delta^{(min)}(x,y)$ into the Mellin integral we can express $h_2(v)$ as a sum over residues. For a function $f(y)$ without extra poles

\begin{eqnarray}
h_2(v) &=& \frac{1}{2\pi i} \int_{-i \infty}^{i \infty} v^{-y}f(y) \Gamma^2(2+y) dy \\
&=& \sum_{n=2}^\infty \frac{v^n \left(f'(-n)-f(-n) \log (v)+2 f(-n) \psi ^{(0)}(n-1)\right)}{\Gamma (n-1)^2}
\end{eqnarray}
where $\psi ^{(0)}(z)$ is the digamma function. In this case
\begin{equation}
f(y)=h^{(0)}_{\delta,0} \frac{2 \left(\delta  (8-3 \delta )+4 y^2\right) \Gamma (2-y)^2}{(\delta -4) \left(\delta ^2-4 y^2\right)}
\end{equation}
so that we need to add the contribution from the pole at $y=-\delta/2$. The relevant sums can be performed with some effort. The final answer is a complicated expression, involving Lerch transcendents, but it admits an expansions around $v=1$:
\begin{eqnarray}
h_2(v) = h^{(0)}_{\delta,0}\sum_{n=0}^\infty  c_n (1-v)^n
\end{eqnarray}
where the general coefficient can be written as follows:
\begin{eqnarray}
c_n = 144 \frac{\Gamma(n+2)}{\Gamma(n+8)} \frac{(18-n)(n^2-1)}{\delta-4} +q_n(\delta) \\
+ \frac{\delta ^3 \left(\delta ^2-4\right)^2  \sin \left(\frac{\pi  \delta }{2}\right) \left(\Gamma \left(-\frac{\delta }{2}\right) \Gamma \left(n+\frac{\delta }{2}\right)+\Gamma \left(\frac{\delta }{2}\right) \Gamma \left(n-\frac{\delta }{2}\right)\right)}{128 \pi  \Gamma (n+1)} \psi ^{(1)}\left(\frac{\delta }{2}\right)
\end{eqnarray}
with $q_n(\delta) $ a polynomial in $\delta$ such that $c_n \sim \frac{1}{\delta}$ for large $\delta$. Note that this fixes the polynomial uniquely. 

\begin{figure}[h!]
\centering
\includegraphics[width=3.5in]{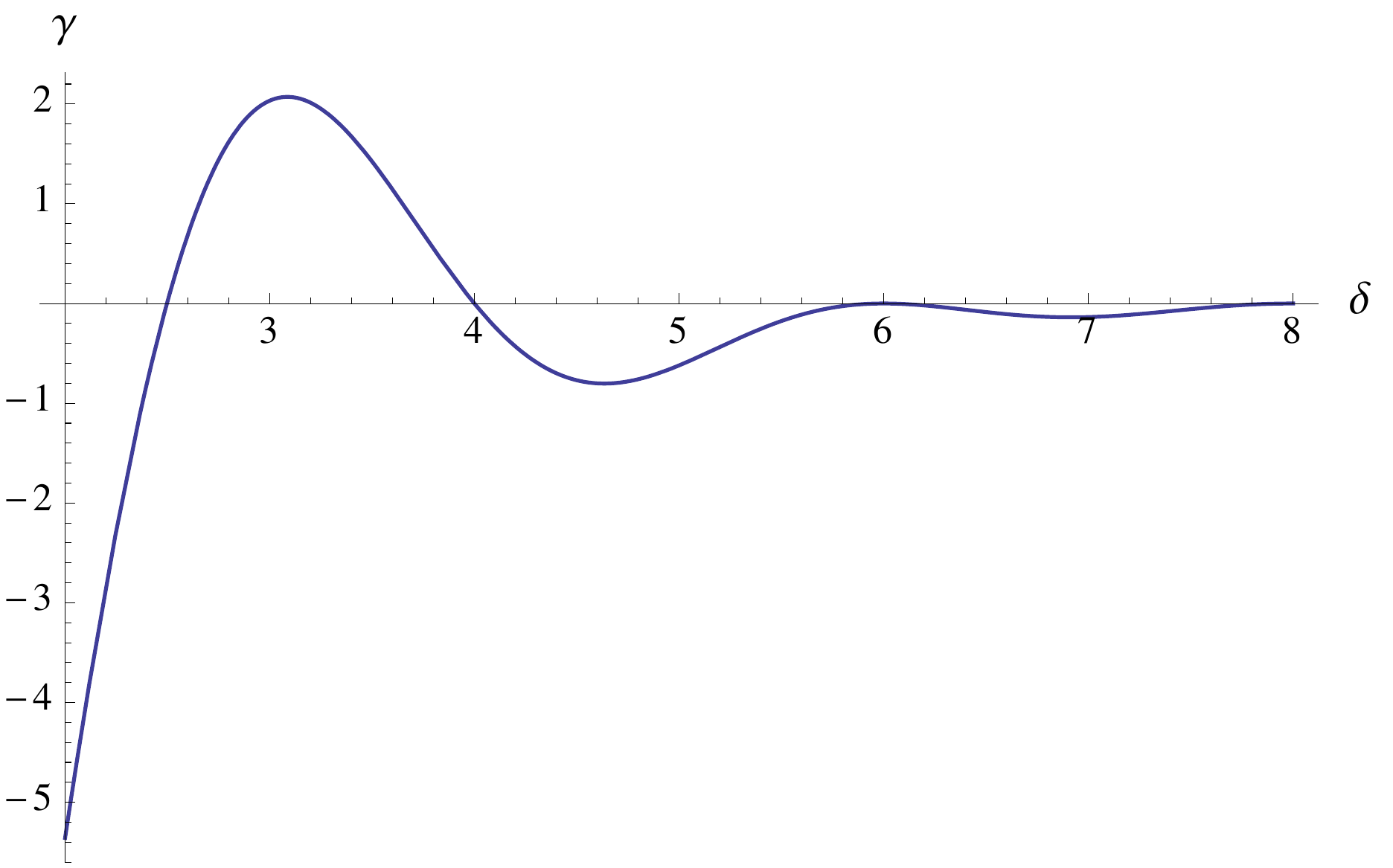}
\caption{Contribution to the anomalous dimension $\gamma_{0,0}$ from the symmetric exchange of a scalar primary, in units of $a_{\delta}$ . \label{gamma00}}
\end{figure}

Having computed the solution in space time we can perform the CPW decomposition:

\begin{equation}
 h_2(v) = \frac{1}{2} \sum_\ell \gamma_{0,\ell} a^{(0)}_{0,\ell} g^{coll}_{4+\ell,\ell}(v)
\end{equation}
where the collinear conformal block $g^{coll}_{4+\ell,\ell}(v)$ is the small $u$ limit of the full block. For instance, for spin zero we find: 

\begin{eqnarray}
\gamma_{0,0} &=& a_{\delta} \left( \left( \frac{5 \delta ^5}{16}+\frac{5 \delta ^4}{16}-\frac{55 \delta ^3}{24}-\frac{5 \delta ^2}{2}+\frac{19 \delta }{6}-\frac{18}{7 (\delta -4)}+5 \right) \frac{\Gamma (\delta +4)}{\Gamma \left(2-\frac{\delta }{2}\right)^2 \Gamma \left(\frac{\delta }{2}+2\right)^4} \right. \nonumber \\
& &  \left. + \frac{5\ 2^{\delta +3} (\cos (\pi  \delta )-1) \Gamma \left(\frac{\delta +5}{2}\right) \psi ^{(1)}\left(\frac{\delta }{2}\right)}{\pi ^{5/2} \Gamma \left(\frac{\delta }{2}+2\right)} \right)
\end{eqnarray}
where we have rewritten the coefficient $h^{(0)}_{\delta,0}$ in terms of the OPE coefficient $a_{\delta}$. It is instructive to study $\gamma_{0,0}$ as a function of $\delta$, see figure \ref{gamma00}.
We observe something very interesting. While the contribution to $\gamma_{0,0}$ can be positive for $\delta<4$, it is always negative for $\delta>4$. In order to draw this conclusion we used the fact that $a_{\delta}$ is positive due to unitarity. 

One can also compute the contributions to $\gamma_{0,\ell}$ for $\ell=2,4,\cdots$, although the results are very lengthy to be reproduced here, and we have not managed to find a closed form expression. From the explicit results one observes that the correction to $\gamma_{0,\ell}$ for $\ell>0$ is always negative. An interesting limit is that of large spin. The simplest way to compute it is along the lines of  \cite{Fitzpatrick:2012yx,Komargodski:2012ek}. Given the analytic structure of the Mellin amplitude  the space time expression contains a term

\begin{equation}
\frac{{\cal A}(u,v)}{v^2} \sim h^{(0)}_{\delta,0} v^{\delta/2-2} u^2 \log u.
\end{equation}
 For $\delta<4$ this gives a divergence, which can be reproduced only with the correct behaviour for $\gamma_{0,\ell}$ at large $\ell$. We obtain
 
\begin{equation}
\gamma_{0,\ell}= -3 a_\delta \frac{\Gamma(4+\delta)}{\Gamma^2\left(2-\frac{\delta}{2} \right)\Gamma^2\left(2+\frac{\delta}{2} \right)} \frac{1}{\ell^\delta} + \cdots
\end{equation}
This precisely agrees with the results of \cite{Fitzpatrick:2012yx,Komargodski:2012ek}, upon specific shifts corresponding to the fact that we are dealing with superconformal blocks. 

As already mentioned, a Mellin amplitude with a finite number of generic poles is not consistent with unitarity. Hence, it is important to extend the computation above in order to include the whole tower of descendants.

\subsubsection{Full and general exchange}
\label{gralexchange}
In the following we will generalise the previous computation to intermediate operators with spin, and including the whole tower of descendants. As a consequence of unitarity, we will see that the sign of the contributions to the anomalous dimensions of twist four operators has a definite sign, as for the simple model above. We propose the following Mellin representation given the exchange of an operator of spin $\ell$ and twist $\tau$:

\begin{equation}
M_\tau^{(\ell)}(x,y) = a_{\tau,\ell} \sum_{k=0,1,\cdots} \alpha^{(\ell)}_k \left( \frac{P_{\ell+\tau}^{(\ell)}(y,z)}{x+\tau/2+k} + \frac{P_{\ell+\tau}^{(\ell)}(x,z)}{y+\tau/2+k} + \frac{P_{\ell+\tau}^{(\ell)}(x,y)}{z+\tau/2+k} \right)+ R_{\ell-1}(x,y,z)
\end{equation}
where $\alpha^{(\ell)}_k$ must be adjusted such as to obtain the correct contribution in space-time:

\begin{equation}
\label{alphak}
\alpha^{(\ell)}_k = \frac{2 (-1)^{k+1} \sin (\pi  \tau ) \Gamma (-k-\tau -2-\ell)}{\Gamma (k+1) \Gamma \left(-k-\frac{\tau }{2}+2\right)^2}
\end{equation}
and $a_{\tau,\ell}$ is the positive OPE coefficient of the single trace operator. $R_{\ell-1}(x,y,z)$ is a completely symmetric polynomial of degree $\ell-1$ which cannot be fixed simply by requiring the correct analytic structure.\footnote{There are several ways to understand the degree of this polynomial. See \cite{Costa:2012cb} for a related discussion.} Note that each of the three pieces in parentheses has the same poles and residues as the conformal block but differs by regular terms.This operation was also considered in \cite{Fitzpatrick:2011dm}, it corresponds to the exchange of a particle in $AdS$ and it leads to a polynomially bounded Mellin expression (while the conformal block is not). In addition, we have the ambiguity of adding a symmetric polynomial $R_{\ell-1}(x,y,z)$ of degree $\ell-1$. In the following discussion we will set this polynomial to zero. We will come back to this issue later, but note that in the case of a scalar exchange there is no such ambiguity.     

The solution can be written as the sum of three pieces (corresponding to exchanges in the $s,t,u$ channels). 

\begin{equation}
M_\tau^{(\ell)}(x,y) = a_{\tau,\ell} \left( Q_\tau^{(\ell)}(x;y,z)+Q_\tau^{(\ell)}(y;z,x)+Q_\tau^{(\ell)}(z;x,y) \right)
\end{equation}
For each spin, the above sum can be performed explicitly. For instance, for the exchange of a scalar operator plus all its tower of descendants we obtain

\begin{equation}
Q_\tau^{(0)}(x;y,z)= \frac{2 \Gamma (\tau +4) \, _3F_2\left(\frac{\tau }{2}-1,\frac{\tau }{2}-1,x+\frac{\tau }{2};x+\frac{\tau }{2}+1,\tau +3;1\right)}{\Gamma \left(\frac{4-\tau }{2}\right)^2 \Gamma \left(\frac{\tau +4}{2}\right)^4 (\tau +2 x)}
\end{equation}
and similar expressions for operators with higher spin. We would like to compute the contribution from such a solution to the anomalous dimension of twist four operators. Finding analytic expressions in this case is harder than before, so that we find it convenient to follow an alternative route. It can be shown, see {\it e.g.} \cite{Goncalves:2014ffa}, that given a Mellin amplitude $M(x,y)$, without poles at $x=-2$, the contribution to the  anomalous dimension of twist four operators is given by

\begin{equation}
 \gamma_{0,\ell}=-\frac{1}{a^{0}_{0,\ell}}\frac{1}{2\pi i} \frac{\sqrt{\pi} \Gamma(\ell+7)}{2^{5+\ell}\Gamma(\ell+7/2)\Gamma(\ell+4)} \int dy  \Gamma^2(y+2) \Gamma^2(2-y) M(-2,y) {\cal F}_\ell(y)
\end{equation}
where the contour of integration runs along the imaginary axis and
\begin{equation}
\label{Hans}
{\cal F}_\ell(y)=\frac{(4_\ell)^2}{\Gamma(\ell+1)} \, _3F_2(-\ell,\ell+7,y+2;4,4;1)
\end{equation}
is a special case of the continuous Hahn polynomial, see \cite{Costa:2012cb}. We can then plug the solutions in Mellin space and compute, numerically, the corresponding contribution to the anomalous dimension of twist four operators of different spin. Figures 2 and 3 show the results for intermediate single trace operators of spin zero and two.

\begin{figure}[h!]
\centering
\includegraphics[width=6.5in]{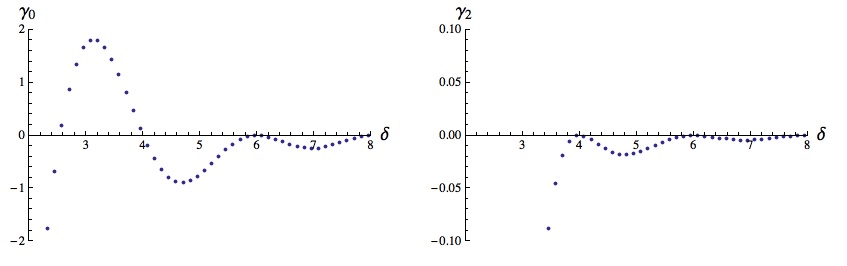}
\caption{Contribution to the anomalous dimension $\gamma_{0,0}$ and $\gamma_{0,2}$ from the exchange of an intermediate operator of spin zero, in units of the positive OPE coefficient. The behaviour of  $\gamma_{0,\ell}$ with $\ell > 0$ is very similar to $\gamma_{0,2}$. \label{spin0}}
\end{figure}

\begin{figure}[h!]
\centering
\includegraphics[width=6.5in]{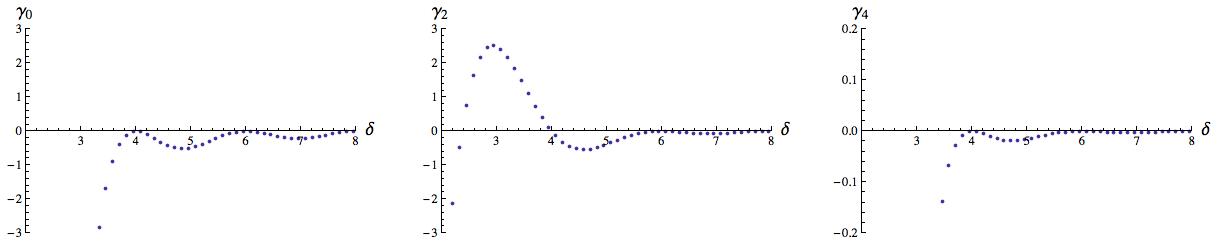}
\caption{Contribution to the anomalous dimension $\gamma_{0,0}$, $\gamma_{0,2}$ and $\gamma_{0,4}$ from the exchange of an intermediate operator of spin two, in units of the positive OPE coefficient. \label{spin2}}
\end{figure}

The results can be summarised as follows. A new single-trace operator of twist higher than four always leads to a {\it negative} contribution to the anomalous dimension of twist-four operators. When exchanging an operator of spin $j$ only the contribution to $\gamma_{0,j}$ can be positive, provided the twist of the operator is smaller than four.

\subsection{Positivity constraints for the Mellin amplitude}
As we have seen above, when considering a new single-trace operator entering at order $1/N^2$ in the CPW decomposition, the contribution to the anomalous dimensions of double trace operators of leading twist has a definite sign. This is related to certain positivity properties of Mack polynomials $P^{(\ell)}_\tau(y,z)$. It is simpler to consider the $s-$channel contribution from the exchange of an operator of twist $\tau$ and spin $\ell$. \footnote{Including the other two channels does not change the sign of the contribution to the anomalous dimension of double trace operators.} In this case, the relevant polynomial is $P^{(\ell)}_\tau(y,-y)$, since only the pole at $x=-2$ contributes. The contribution to the  anomalous dimension $\gamma_{0,j}$ is proportional  to

\begin{equation}
\gamma_{0,j} \sim \int dy \Gamma^2(y+2) \Gamma^2(2-y) {\cal F}_j(y) P_{\ell+\tau}^{(\ell)}(y,-y) 
\end{equation}
where the polynomials ${\cal F}_j(y)$ have been introduced in (\ref{Hans}). It can be explicitly checked that for $\tau=4$ the Mack polynomials $P_{\ell+\tau}^{(\ell)}(y,-y)$ reduce precisely to ${\cal F}_j(y)$, up to a proportionality factor. For $\tau \geq 4$ the following positivity condition can be verified:

\begin{eqnarray}
\int dy \Gamma^2(y+2) \Gamma^2(2-y)  {\cal F}_j(y) P_{\ell+\tau}^{(\ell)}(y,-y) \geq 0,~~~~\mathrm{for}~\tau \geq 4
\end{eqnarray}
while the equality is satisfied for $\tau=4$ and $j<\ell$. On the other hand, the above integral is identically zero for $j>\ell$. An equivalent formulation of the above positivity condition is the statement that Mack polynomials $P_{\ell+\tau}^{(\ell)}(y,-y)$ admit a decomposition:

\begin{equation}
P_{\ell+\tau}^{(\ell)}(y,-y) = \sum_{j=0}^\ell c_j(\tau) {\cal F}_j(y)
\end{equation}
where $c_j(\tau) \geq 0$ for $\tau \geq 4$. Let us assume we are in a regime in which the twist of all single trace unprotected operators is higher than four. The Mellin expression is a meromorphic function with single poles corresponding to exchanged operators. The above discussion suggests that the Mellin transform admits an analogous decomposition. More precisely:

\begin{equation}
M(-2,y,-y) = \sum_{n=0,2,\cdots} c_n {\cal F}_n(y)
\end{equation}
where $c_n \geq 0$. This would lead to a negative anomalous dimension for all twist four double trace operators. Since the Mellin amplitude is symmetric under exchange of the Mellin variables, this can also be written as

\begin{equation}
\label{positivity}
M(x,-2,-x) = \sum_{n=0,2,\cdots} c_n {\cal F}_n(x),~~~~~~c_n \geq 0.
\end{equation}
For instance, one can explicitly check that this holds for the supergravity term. An important comment is in order. The relations that have led to (\ref{positivity}) have been shown for the case of a scalar exchange. In the case of intermediate operators with spin we have set certain ambiguous polynomial to zero, which could, in principle spoil the positivity properties.\footnote{Note however, that the contribution to the anomalous dimension of twist four operators of spin higher or equal than that of the exchanged operator will not be affected.}. Assuming this does not happen, we can propose a slightly stronger result. Let us call $\tau_\ell^{min}$ the minimum twist for single-trace operators of spin $\ell$. If we are in a regime in which $\tau_\ell^{min}>4$ for all $\ell \geq \ell^*$, then (\ref{positivity}) holds with $c_{\ell^*},c_{\ell^*+1},\cdots$ positive. This will lead to negative $\gamma_{0,\ell}^{extra}$ for $\ell \geq \ell^*$. In the following we will argue that the same positivity condition (\ref{positivity}), together with its stronger version, follows from different considerations. 

\subsection{Positivity property from Regge behaviour}
In all the examples studied so far we have verified the positivity condition (\ref{positivity}) for the Mellin amplitude. Let us see how this condition arises under general mild assumptions \footnote{We would like to thank J. Penedones for his suggestion to look into the Regge behaviour in relation with this problem.}. Assume the Mellin amplitude $M(x,y,z)$ displays Regge behaviour for fixed $y$ and large $x$. More precisely, let us assume that for a fixed large enough $y$, $M(x,y,z)$ goes to zero for large $x$. It then follows that $M(x,y,z)$ for that value of $y$ is given by a sum over poles with the correct residues. 

\begin{equation}
\label{regge}
M(x,y,-2-y-x) = \sum_{\tau,\ell} a_{\tau,\ell}\sum_k \alpha_k^{(\ell)} \left( \frac{P_{\ell+\tau}^{(\ell)}(y,-2-y+\tau/2)}{x+\tau/2+k} + \frac{P_{\ell+\tau}^{(\ell)}(-2-y+\tau/2,y)}{-2-x-y+\tau/2+k} \right)
\end{equation}
where the sum runs over intermediate single-trace operators (we have subtracted the supergravity contribution), with positive OPE coefficients $a_{\tau,\ell}$. Note that this expression does not have any ambiguities. Next, we ask how large does $y$ have to be. From the discussion above, it follows that the first pole for $y$ is at $y=-\tau^{min}/2$, where $\tau^{min}$ is the twist of the single-trace operator with minimal twist. Let us first assume we are in a regime in which $\tau^{min} > 4$. Then (\ref{regge}) should also hold for $y=-2$ \footnote{The Regge trajectory is defined by $\tau^{min}_\ell$, so that this behaviour holds for $y>-\tau^{min}_0/2$ in our conventions.}. In this case we can write

\begin{equation}
M(x,-2,-x) = \sum_{\tau,\ell} a_{\tau,\ell}\sum_k \alpha_k^{(\ell)} \left( \frac{P_{\ell+\tau}^{(\ell)}(-2,\tau/2)}{x+\tau/2+k} + \frac{P_{\ell+\tau}^{(\ell)}(\tau/2,-2)}{-x+\tau/2+k} \right)
\end{equation}
One can explicitly check that the combination $\alpha_k^{(\ell)}P_{\ell+\tau}^{(\ell)}(-2,\tau/2)$ is non-negative. Furthermore, the coefficients $a_{\tau,\ell}$ are positive due to unitarity. Finally, the combination

\begin{equation}
 \frac{1}{x+\tau/2+k} + \frac{1}{-x+\tau/2+k} 
\end{equation}
admits a decomposition in terms of continuous Hahn polynomials with positive coefficients. Assuming that the spectrum is such that the sum over $\tau$ and $\ell$ converges, the positivity condition (\ref{positivity}) then follows. The stronger version of the positivity condition can be understood as follows. Let us assume we are in a regime in which  $\tau^{min}_\ell > 4$ for $\ell=2,4,\cdots$ but $\tau^{min}_0<4$. In this case, the Mellin amplitude at $y=-2$ will not fall at infinity, but tend to a constant. This extra constant term may affect the sign of $c_0$ in (\ref{positivity}), but it will not affect the others. If we now assume that also $\tau^{min}_2<4$, then we will have an extra constant plus a quadratic term in $x$ at infinity. This may affect the sign of $c_0,c_2$ but not the others. And so on.

\subsubsection{Consequences for the spectrum of  ${\cal N}=4$ SYM }
Let us focus our attention in twist four double trace operators. Their scaling dimension at order $1/N^2$ is given by

\begin{equation}
\Delta_{0,\ell} = 4+ \ell - \frac{1}{N^2} \frac{96}{(\ell+1)(\ell+6)} + \frac{1}{N^2}  \gamma_{0,\ell}^{extra}
\end{equation}
where we have included the supergravity result and $\gamma_{0,\ell}^{extra}$ corresponds to the contribution from the rest of the solution to the crossing relations. In a regime in which the twist of all single-trace operators is higher than four, we obtained the positivity condition (\ref{positivity}). This will lead to negative $\gamma_{0,\ell}^{extra}$. Furthermore, in a regime in which $\tau_\ell^{min}>4$ for all $\ell \geq \ell^*$, then (\ref{positivity}) holds with $c_{\ell^*},c_{\ell^*+1},\cdots$ positive. This will lead to negative $\gamma_{0,\ell}^{extra}$ for $\ell \geq \ell^*$. 

What are the consequences for the spectrum of leading twist operators? Let us focus in a given spin.  If $\tau_\ell^{min}$ is smaller than four, then this will be the minimal twist for that spin. If, on the other hand,  $\tau_\ell^{min}>4$, then the minimal twist will be given by $\Delta_{0,\ell}-\ell$, with $\Delta_{0,\ell}$ given above. Since we have argued that $\gamma_{0,\ell}^{extra}$ is always negative, then $4- \frac{1}{N^2} \frac{96}{(\ell+1)(\ell+6)}$ provides an upper bound for the minimal twist. This exactly agrees with the numerical bounds observed in \cite{Beem:2013qxa}. 

\section{Limiting cases}

\subsection{Operators with large dimension}
In order to make contact with the results of \cite{Heemskerk:2009pn,Alday:2014tsa} we would like to consider a limit in which the dimension of the unprotected single trace operators becomes very large. In ${\cal N}=4$ SYM this happens for large $\lambda$. Hence, we would like to consider the solutions constructed in the previous section, and study them as $\tau$ becomes very large, for finite values of the Mellin variables $(x,y,z)$. This limit is very similar to the flat space limit studied in \cite{Penedones:2010ue}, although in that limit the Mellin variables are also large. As in that case, the limit is a bit subtle, and one is to perform the sum over descendants before taking the limit. Let us consider the partial terms

\begin{equation}
Q_\tau^{(\ell)}(x;y,z) = \sum_{k=0,1,\cdots} \alpha^{(\ell)}_k \frac{P_{\ell+\tau}^{(\ell)}(y,z)}{x+\tau/2+k} 
\end{equation}
In order to proceed, we note that for large $\tau$, the leading contribution comes from the region $k=\xi \tau^2$, with finite $\xi$. In that limit the sum over $k$ becomes a integral:

\begin{equation}
\sum_k \to \tau^2 \int_0^\infty d\xi
\end{equation}
Given the explicit expression for $\alpha^{(\ell)}_k $ in (\ref{alphak}) we can perform the expansion.\footnote{It is convenient to transform $\alpha^{(\ell)}_k$ into an equivalent expression, where $\Gamma-$functions contain only $\tau$ with positive sign. This can be easily done with the help of the Euler's reflection formula $\Gamma(1-z)\Gamma(z) \sin(\pi z) = \pi$.} We obtain

\begin{equation}
Q_\tau^{(\ell)}(x;y,z) =\frac{4 (\cos(\pi \tau)-1)}{\pi} \frac{P_{\ell+\tau}^{(\ell)}(y,z)}{\tau^{4\ell}} \int_0^\infty \frac{e^{-\frac{1}{4 \xi }}}{\xi^{2\ell+7}}\left(-\frac{ 1}{4 \tau ^{12}} +\frac{(4 (4 \ell+9) \xi -1)}{32 \xi^2 \tau ^{13}}+ \cdots \right) d \xi
\end{equation}
The integral over $\xi$ is convergent and can be performed order by order in $1/\tau$. The large $\tau$ behaviour of Mack polynomials is given by 

\begin{equation}
P_{\ell+\tau}^{(\ell)}(y,z) = \frac{16}{\pi^2} 4^\tau \tau^{\ell} + \cdots,
\end{equation}
Let us focus in the scalar exchange, with $\ell=0$. At each order in $1/\tau$ the final result for the symmetrized solution is a polynomial in the symmetric variables:

\begin{eqnarray}
\sigma_2 = x^2+y^2+z^2\\
\sigma_3 = x^3+y^3+z^3
\end{eqnarray}
In terms of these the large $\tau$ expansion takes the following form

\begin{equation}
\label{largedelta}
M_\tau^{(0)}(x,y) = \frac{\hat a_\tau}{\tau^{12}} \left(\left(1+ \cdots \right) + \frac{\sigma_2}{\tau^4} \left(224 + \cdots  \right) + \frac{\sigma_3}{\tau^6} \left(-7168 + \cdots \right)  + \cdots \right)
\end{equation}
where we have introduced 

\begin{equation}
\label{ahat}
\hat a_\tau = \frac{5\, 3^2\, 2^{14} \left(1-\cos (\pi  \tau )\right) \Gamma (\tau +3) \Gamma (\tau +4)}{\pi ^2 \Gamma^4 \left(\frac{\tau }{2}+2\right)} a_{\tau,0}
\end{equation}
In the picture of \cite{Heemskerk:2009pn,Alday:2014tsa} it was assumed that single trace operators acquire a very large twist $\tau \sim \Delta_{gap}$. In this case the relative coefficients in the expansion (\ref{largedelta}) exactly agree with the ones in \cite{Alday:2014tsa}. In that paper this behaviour was argued from the point of view of effective field theories and dimensional analysis, following the discussion in \cite{Fitzpatrick:2010zm, Fitzpatrick:2012cg}.

Let us now focus in the overall coefficient $\hat a_\tau$. First note that the prefactor in the definition of (\ref{ahat}) grows exponentially:
\begin{equation}
 \frac{\Gamma (\tau +3) \Gamma (\tau +4)}{\Gamma^4 \left(\frac{\tau }{2}+2\right)} \sim 2^{2\tau}
\end{equation}
We expect $a_{\tau,0}$ to decay exponentially for large $\tau$, so that $\hat a_\tau$ has a power law behaviour. Indeed, it has been argued in \cite{Pappadopulo:2012jk} that convergence of the OPE puts bounds on the behaviour of OPE coefficients at large dimensions. While it would be interesting to understand the behaviour of  $\hat a_\tau$ for large $\tau$ in general, one can analyse the problem at tree-level. In that case $\tau = 4,6,\cdots$ and the OPE coefficients behave as:

\begin{equation}
a_{\tau,\ell}^{(0)} \sim 2^{-2\tau} 
\end{equation}
with the expected exponential behaviour. Furthermore, in \cite{Costa:2012cb, Minahan:2014usa}, it was argued that the structure constant for two protected operators of dimension two and one unprotected operator of large twist $\tau$ at large $N$ is given by:

\begin{equation}
\label{OPElargetau}
a_{\tau,0} \sim \lambda^{3/2} 2^{-2\tau} \csc^2(\pi \tau/2)
\end{equation}
In addition to the expected exponential behaviour, note that the poles present in the factor $\csc^2(\pi \tau/2)$ cancel neatly against the zeroes in $\hat a_\tau$. In the limit of large dimension we obtain

\begin{equation}
 \frac{\hat a_\tau}{\tau^{12}}  \sim \frac{ \lambda^{3/2}}{\tau^{12}}
\end{equation}
For large $\lambda$ we have $\tau = \Delta_{gap} \sim \lambda^{1/4}$, so that 

\begin{equation}
 \frac{\hat a_\tau}{\tau^{12}}  \sim \lambda^{-3/2}
\end{equation}
which is exactly the expected result! In order to recover the results of our previous paper, it would be interesting to prove the scaling (\ref{OPElargetau}) from a purely CFT perspective. In any case, our analysis suggests that the polynomial solutions considered in \cite{Alday:2014tsa} may be seen as remnants of the non-polynomial solutions corresponding to single trace operators, when the operators become very massive. Finally, note that from this analysis the sign in front of the above series is fixed, and such as to ensure that the contribution to the anomalous dimension of leading twist operators (in this case four) is always negative. The same analysis can be carried out for intermediate operators with spin. Disregarding the ambiguous polynomial $R_{\ell-1}(x,y,z)$, the large twist behaviour is exactly as expected, with the correct signs as to give a negative contribution to the anomalous dimension of leading twist operators. 

\subsection{Flat space limit and relation to causality}

There is a very simple relation between the Mellin representation for a large $N$ CFT correlator and the S-matrix of the bulk dual theory in the flat space limit \cite{Penedones:2010ue}. Given the Mellin amplitude $M(x,y,z)$ the flat space S-matrix in our conventions is given by

\begin{equation}
\label{flatspace}
{\cal T}(s,t,u) =-2\lim_{\lambda \to \infty} \lambda^{3/2}  \oint \frac{d \alpha}{2\pi i} \frac{e^{-\alpha} }{ \, \alpha^6} M\left(\frac{\sqrt{\lambda} s}{\alpha} ,\frac{\sqrt{\lambda} t}{\alpha}  ,\frac{\sqrt{\lambda} u}{\alpha} \right)
\end{equation}
For instance, the supergravity term leads to the following contribution

\begin{equation}
{\cal T}_{sugra}(s,t,u) = \frac{16}{s t u}.
\end{equation}
Next, we can consider the flat space limit of an $s-$channel exchange:

\begin{equation}
\left.M_{\tau}^{(\ell)}(x,y,z)\right|_{s-channel} = \sum_{k=0,1,\cdots} \alpha_k^{(\ell)} \frac{P_{\ell+\tau}^{(\ell)}(y,z)}{x+\tau/2+k}
\end{equation}
As $\lambda$ becomes large the twist of the single trace operator scales as $\tau \sim \lambda^{1/4}$ so that the flat space limit corresponds to large $\tau$ with the Mellin variables scaling as $x,y,z \sim \tau^2$. In this limit the Mack polynomials reduce to the Harmonic functions on $S^3$ and we obtain

\begin{eqnarray}
\left.M\left(\frac{\sqrt{\lambda} s}{\alpha} ,\frac{\sqrt{\lambda} t}{\alpha}  ,\frac{\sqrt{\lambda} u}{\alpha} \right) \right|_{s-channel}\sim \frac{\sin((\ell+1)\theta)}{\sin \theta} \times\\
\left(\frac{\sqrt{\lambda}}{\tau^2 \alpha} \right)^\ell \left(-\Gamma(6+\ell) +4 \Gamma(7+\ell) \frac{\sqrt{\lambda} s}{\tau^2 \alpha} - 16 \Gamma(8+\ell) \left(  \frac{\sqrt{\lambda} s}{\tau^2 \alpha} \right)^2 +\cdots \right)
\end{eqnarray} 
The first factor arises from the corresponding limit of the Mack polynomial. We have introduced the scattering angle $t/u=\frac{1-\cos \theta}{1+\cos \theta}$. Performing the contour integral we obtain
\begin{equation}
{\cal T}(s,t,u)  \sim \frac{\sin((\ell+1)\theta)}{\sin \theta} \frac{1}{s+m^2}
\end{equation}
 where $m^2=\tau^2/(4 \sqrt{\lambda})$. This is the correct flat space propagator for a particle of mass $m$ and spin $\ell$. Notice that the overall coefficient we have suppressed is {\it positive}. The precise flat space limit of the non-polynomial solutions will depend, in general, on the polynomials $R_{\ell-1}(x,y,z)$ \footnote{Although the degree of this polynomial is smaller than $\ell$, the overall coefficients may scale with $\tau$.}. The full Mellin amplitude at order $1/N^2$ should be such that its flat space limit reproduces the Virasoro Shapiro amplitude: 
  \begin{equation}
  {\cal T}(s,t,u)={\cal T}_{VS}(s,t,u)= \frac{16}{s t u} \frac{\Gamma(1-s/4)}{\Gamma(1+s/4)} \frac{\Gamma(1-t/4)}{\Gamma(1+t/4)} \frac{\Gamma(1-u/4)}{\Gamma(1+u/4)}
 \end{equation}
 where recall $s+t+u=0$. From our discussion in section 3.1, we expect that in the large $\lambda$ limit the Mellin expression reduces to the supergravity term plus a series of completely symmetric polynomial terms suppressed by powers of $1/\lambda$:
 
 \begin{equation}
 M(x,y,z) = -\frac{16}{(x+1)(y+1)(z+1)} +\frac{p^{(0)}}{\lambda^{3/2}}+\frac{p^{(2)}(x,y,z)}{\lambda^{5/2}}  + \cdots 
 \end{equation}
This expansion, with the appropriate signs, is consistent with the Virasoro Shapiro amplitude. Finally, let us add that the proper flat space limit fixes the leading term of each polynomial $p^{(i)}(x,y,z)$, see \cite{Goncalves:2014ffa} where this exercise was performed. It would be interesting to see what else can be said about sub-leading terms. 

One may ask whether the polynomials $R_{\ell-1}(x,y,z)$ may be set to zero and we can obtain the full Mellin transform at order $1/N^2$ as the sum of the non-polynomial solutions  presented above. It turns out that these polynomials are necessary for consistency with the flat space limit.\footnote{J. Penedones, private communication.}

\subsubsection*{Relation to causality}

Causality constraints on effective field theories were studied in \cite{Adams:2006sv}. It was determined that the S-matrix of a low energy effective field theory should satisfy certain positivity constraints if the theory has a consistent UV completion. More precisely, in the forward limit $t \to 0$ the regular part of the S-matrix has an expansion

\begin{equation}
  {\cal T}(s,0,-s) = \alpha + \beta s^2 +\gamma s^4+ \cdots
\end{equation}
where all the coefficients $\alpha,\beta,\cdots$ are non-negative. This can be checked for the flat space limit of the solutions considered above. Indeed, in the forward limit $\theta \to 0$ the corresponding S-matrix reduces to

\begin{equation}
{\cal T}(s,0,-s) \sim  \frac{1}{s+m^2}+\frac{1}{-s+m^2} + \frac{1}{m^2} =\frac{3}{m^2} + \frac{2 s^2}{m^6}+ \cdots
\end{equation}
with only positive coefficients. Hence, the flat space limit of the non-polynomial solutions constructed above, assuming $R_{\ell-1}(x,y,z)=0$, is consistent with the causality constraints of \cite{Adams:2006sv}.\footnote{The reason for this is obvious, since in the flat space limit we simply recover a sum of propagators in the three channels, with positive coefficient.} The same is true for the Virasoro-Shapiro amplitude. 

Let us now consider the flat space limit, eq. (\ref{flatspace}), in the forward limit, for a general Mellin amplitude. After substracting the supergravity contribution (which is not regular in the forward limit), the relevant Mellin amplitude will have an expansion:

\begin{equation}
 M\left(\frac{\sqrt{\lambda} s}{\alpha} ,0  ,-\frac{\sqrt{\lambda} s}{\alpha} \right) = c_0 + c_1 \frac{\lambda s^2}{\alpha^2} + c_2 \frac{\lambda^2 s^4}{\alpha^4} +\cdots
\end{equation}
where the coefficients $c_i$ depend on $\lambda$, and only the leading term survives in the flat space limit. Consistency with causality then implies that all the coefficients $c_i$ are positive at leading order in $\lambda$. This condition has exactly the same form as (\ref{positivity})! Actually, it follows from (\ref{positivity}): The Hahn polynomials have positive coefficients, and in the flat space limit only the leading term from each polynomial will survive. On the other hand (\ref{positivity}) is also valid for finite (but sufficiently large) $\lambda$.  

\subsection{Large $n$ limit}
Before concluding, let us make a few remarks on the large $n$ behaviour of $\gamma_{n,\ell}$. We are interested in the limit $n \gg 1$ for finite $\ell$. This behaviour is controlled by the same harmonic functions which arise in the flat space limit. Given the Mellin representation $M(x,y,z)$ the leading large $n$ behaviour of the anomalous dimension $\gamma_{n,\ell}$ is given by

\begin{equation}
\label{largen}
\gamma_{n,\ell} = \frac{1}{n(\ell+1)} \oint \frac{d \alpha}{2\pi i} \frac{e^{-\frac{n^2}{2}\alpha} }{ \, \alpha^6} \int_0^{2\pi} \frac{\sin^2\theta}{\pi} P_\ell(\theta) M(\frac{2}{\alpha}, -\frac{1-\cos \theta}{\alpha}, -\frac{1+\cos \theta}{\alpha}) d\theta
\end{equation}
This allows to compute the large $n$ behaviour due to different terms. For instance, from the supergravity contribution we get the following dependence:

\begin{equation}
 \gamma_{n,\ell}^{sugra} = -2\frac{n^3}{1+\ell} + \cdots
\end{equation}
which can also be computed from the explicit answer. For the non-polynomial solutions corresponding to the exchange of an operator of spin $\ell$ we obtain:

\begin{equation}
 \gamma_{n,\ell}^{\tau,0} \sim - n^{7+2\ell}
\end{equation}
where we have stressed the fact that the overall coefficient is negative and the twist of the operator has been kept fixed in the limit. Note that the limit is insensitive to the polynomials $R_{\ell-1}(x,y,z)$, but the addition of a higher order polynomial would spoil this behaviour. Furthermore, note that the supergravity term is the solution consistent with crossing symmetry and the structure of the CPW decomposition that leads to the smallest growing with $n$ for $\gamma_{n,\ell}$.

Having the expression (\ref{largen}) for the large $n$ behaviour of $\gamma_{n,\ell}$ in terms of an arbitrary Mellin amplitude, we would like to compute the leading large $n$ behaviour to all order in $1/\sqrt{\lambda}$, in ${\cal N}=4$ SYM. As already mentioned, the Mellin amplitude should be such that its flat space limit reproduces the Virasoro Shapiro amplitude. On the other hand, note that the integral expression giving the flat space limit (\ref{flatspace}) is almost the same expression which gives the leading large $n$ contribution (\ref{largen}). Hence, the leading large $n$ behaviour can be directly written in terms of the Shapiro Virasoro amplitude! We obtain
 
 \begin{equation}
 \gamma_{n,\ell} = \frac{\lambda^{-3/2} n^9}{64(\ell+1)}  \int_0^{2\pi} \frac{\sin^2\theta}{\pi} P_\ell(\theta) {\cal T}_{VS}\left( \frac{n^2}{\sqrt{\lambda}}, -\frac{1}{2} \frac{n^2}{\sqrt{\lambda}}(1-\cos \theta),  -\frac{1}{2} \frac{n^2}{\sqrt{\lambda}}(1+\cos \theta) \right) d\theta
 \end{equation}
 which is valid to all orders in $1/\sqrt{\lambda}$. For instance, in the large $\lambda$ limit only the leading term contributes and we obtain
 
  \begin{equation}
 \gamma_{n,\ell}^{sugra} =- \frac{n^3}{4(\ell+1)}   \int_0^{2\pi} \frac{\sin^2\theta}{\pi} P_\ell(\theta) \frac{4}{1-\cos^2 \theta} d\theta = -\frac{2n^3}{\ell+1}
 \end{equation}
 which agrees with the large $n$ behaviour of the supergravity result. For finite $\frac{n^2}{\sqrt{\lambda}}$ we see this expression has poles at $\frac{n^2}{\sqrt{\lambda}} = 4, 8, \cdots$. Note that at these values the dimension $2\Delta+2n$ agrees with twice the dimension of the Konishi operator. We expect these poles to be an artifact of the large $N$ expansion and the fact that we are considering only the leading contribution at each order in $1/\sqrt{\lambda}$. In any case, it is instructive to study the above expression in the limit of large $\frac{n^2}{\sqrt{\lambda}}$. We can avoid the poles by adding a small imaginary part to $\frac{n^2}{\sqrt{\lambda}}$. For large $\frac{n^2}{\sqrt{\lambda}}$ the integral over $\theta$ receives most of its contribution from the saddle points at $\theta=0$ and $\theta=\pi$. The final result is a slower growing with $n$ than the supergravity result. More precisely we obtain $\gamma_{n,\ell} \sim n^2/\log^{1/2} n$. It would be interesting to understand this result further.

 \subsubsection*{Lorentzian singularity}
A growth in the anomalous dimensions of double trace operators, $\gamma_{n,\ell} \sim n^\kappa$, will generically lead to a singularity in the Lorentzian correlator at the point $z =\bar z$. Such singularities are a diagnostic of bulk locality \cite{Gary:2009ae, Heemskerk:2009pn, Maldacena:2015iua} . More precisely we introduce $z=\sigma e^\rho$ and $\bar z = \sigma e^{-\rho}$ and take the limit $\rho \to 0$. The Euclidean correlator is not singular in this limit. However, if we analytically continue to the Lorentzian regime, as described in \cite{Heemskerk:2009pn}, and then take the limit $\rho \to 0$, a singularity arises. The simplest example is that of a single conformal block. Let us consider 

\begin{equation} 
g_{2,0}(z,\bar z) = \frac{\log(1-\bar z)-\log(1-z)}{z-\bar z}
\end{equation}
clearly, this does not have a singularity at $z=\bar z$. The analytic continuation to the Lorentzian regime is explained for instance in appendix B to \cite{Gary:2009ae}. The strategy is to place all the branch cuts of a given expression along the positive real axis (which is already done for the expression at hand). Then along the analytic continuation $z$ does not cross any branch cuts while $\bar z$ crosses all of them. In the example above this leads to

\begin{eqnarray}
\log(1-z) &\to& \log(1-z)\\
\log(1-\bar z) &\to& \log(1-\bar z) + 2\pi i
\end{eqnarray}
so that after the analytic continuation we get a divergence $\frac{2\pi i}{z - \bar z} \sim \rho^{-1}$. In addition to the divergence for each conformal block there is an enhancement effect arising from the large $n$ behaviour of $\gamma_{n,\ell}$. This can be seen as follows. Under the analytic continuation each conformal block acquires an extra phase:

\begin{equation}
u^{\frac{\Delta-\ell}{2}}g_{\Delta+4,\ell}(u,v) \to e^{-i \pi \Delta} u^{\frac{\Delta-\ell}{2}}g_{\Delta+4,\ell}(u,v) 
\end{equation}
when summing over double trace operators all phases add up, since $e^{-2\pi i n}=1$. Furthermore $e^{-i \pi \Delta_{n,\ell}} \sim -i \pi \frac{\gamma_{n,\ell}}{N^2}$. The extra factor $\gamma_{n,\ell}$ results in a enhanced divergence when summing over $n$.  The final divergence for the correlator takes the form\footnote{The translation of the results of \cite{Heemskerk:2009pn} to the case of superconformal blocks is straightforward, and the divergence has exactly the same form.} 

\begin{equation}
{\cal A}(z,\bar z) \sim \sum_{n,\ell}  \frac{n^2 \gamma_{n,\ell}}{2\rho~\sin^2\frac{\theta}{2}} e^{-2i n \rho \tan\frac{\theta}{2}} P_\ell(\theta) 
 \end{equation}
as $\rho$ approaches zero. We have introduced $\sigma = \sin^2\frac{\theta}{2}$. The sum over $n$ generically enhances the divergence. The leading divergence can be computed by approximating the sum by an integral:

\begin{equation}
\sum_n n^\alpha e^{-\beta n} \sim \int_0^\infty dn n^\alpha e^{-\beta n} = \frac{\Gamma(1+\alpha)}{\beta^{\alpha +1}}
\end{equation}
For instance, the supergravity contribution will lead to a divergence $\rho^{-7}$ while the non-polynomial solutions corresponding to a scalar exchange will lead to a divergence $\rho^{-11}$ and so on. Note that the divergences are much more severe in ${\cal N}=4$ SYM than in standard large $N$ CFT. 

Given the leading large $n$ behaviour of the anomalous dimensions of double trace operators to all orders in $1/\sqrt{\lambda}$, eq. (\ref{largen}), we can write down the leading bulk-point divergence to all orders in $1/\sqrt{\lambda}$. This can be writen in terms of the Virasoro-Shapiro amplitude and is given by

\begin{equation}
\label{N4sing}
\left. {\cal A}(z,\bar z) \right|_{div} \sim \sum_n  \frac{2 \lambda^{-3/2} n^{11} }{\rho~\sin^2\frac{\theta}{2}} {\cal T}_{VS}\left( \frac{n^2}{\sqrt{\lambda}}, -\frac{1}{2} \frac{n^2}{\sqrt{\lambda}}(1-\cos \theta),  -\frac{1}{2} \frac{n^2}{\sqrt{\lambda}}(1+\cos \theta) \right) e^{-2i n \rho \tan\frac{\theta}{2}}
\end{equation}
This expression is not unexpected, since in general large N CFT's the residue at the singularity should be related by the flat space S-matrix \cite{Heemskerk:2009pn}. Let us make the following interesting remark. The divergence as $\rho \to 0$ is controlled by the terms with large $n$. For generic angles this is controlled by the Virasoro-Shapiro amplitude at large momentum transfer and fixed angle. In this regime the Virasoro-Shapiro amplitude is known to decay exponentially. As a result, we do not get an enhanced divergence respect to the divergence of each individual conformal blocks. This agrees with the expectation of \cite{Maldacena:2015iua}. Since $\gamma_{n,\ell}$ does grow with $n$ (see discussion above) for large $n^2/\sqrt{\lambda}$, this seems confusing. What happens is that for a generic angle all spins contribute and their divergences cancel out. This is somewhat similar to the chaotic phenomenon mentioned in \cite{Maldacena:2015iua}. 

\section{Conclusions}

In this paper we have considered the four-point correlator of the stress tensor multiplet in ${\cal N}=4$ SYM. The contribution from intermediate operators in shortened supermultiplets can be resummed exacly. As a result, the correlator can be written in terms of a single non-trivial function ${\cal G}(u,v)$, which receives contributions from unprotected operators only. We have analysed the crossing relations in the large $N$ limit, to order $1/N^2$. In order to do so it is convenient to work in Mellin space. The prefactor in the definition of the Mellin amplitude authomatically includes the poles corresponding to double trace operators. At large t'Hooft coupling $\lambda$ the dimension of single trace operators is parametrically large and the Mellin amplitude does not contain extra poles. The solutions reduce to the polynomial solutions previously found. 

In this paper we have focused in a regime in which the dimension of single trace operators is finite. As a consequence they enter, at order $1/N^2$, as intermediate states in the OPE of two external operators. This leads to a structure of simple poles in the Mellin amplitude, and we have considered solutions consistent with crossing and these analytic properties. An important point is that the overall sign of these solutions is fixed by unitarity. We then studied the contribution from such non-polynomial solutions to the anomalous dimension of twist four operators, and have shown that the contribution is always ${\it negative}$, provided the twist of the intermediate operators is larger than four. This follows from certain positivity properties of Mack polynomials. This can be extended to a positivity constraint for a slice of the Mellin amplitude (at order $1/N^2$). In doing so, we have disregarded a polynomial ambiguity that arises when the exchanged particles are not scalar. However, we have shown that the same positivity constraint follows from requiring the correct Regge behaviour for the Mellin amplitude. This positivity condition explains results observed from the numerical bootstrap.

It is instructive to consider the large twist limit of the non-polynomial solutions. On one hand, it reproduces the series of polynomial solutions considered in \cite{Heemskerk:2009pn,Alday:2014tsa}, with the correct suppression factors of an effective field theory. On the other hand, it "predicts" specific overall signs, which in the effective theory treatment were arbitrary. These signs are such that the correction to the anomalous dimensions of twist four operators is negative. For intermediate scalar operators the analysis can be done explicitly, while in the case of intermediate operators with spin, one would need to assume that the polynomial terms do not spoil these properties. 

It is also instructive to consider the limit in which the Mellin amplitude reduces to the S-matrix of the bulk dual theory in flat space, proposed by Penedones in \cite{Penedones:2010ue}. We have analysed this in the forward limit and found a connection between the positivity property for the Mellin amplitude mentioned above, and positivity constrains derived from causality in \cite{Adams:2006sv}. Furthermore, in this limit one should recover the full Virasoro Shapiro amplitude. This has allowed us to write down the large $n$ behaviour of $\gamma_{n,\ell}$, to all orders in $1/\sqrt{\lambda}$ and the leading order bulk-point singularity, to all orders in $1/\sqrt{\lambda}$. We have observed that for finite $\lambda$ the singularity of single conformal blocks is not enhanced. 

There are several open problems that would be interesting to address. It would be interesting to obtain similar positivity constraints for $\gamma_{n,\ell}$ for generic $n$. It would be interesting to understand better how to fix the polynomial ambiguities for the case of the exchange of a single-trace operator with spin. Even assuming polynomial boundedness of the Mellin amplitude, one can always add regular, polynomial terms without compromising the analytic structure of the solutions. As already mentioned, consistency with the flat space limit does require these polynomials. On the other hand, the positivity condition on the Mellin transform was proven regardless of this ambiguity. 

It would also be desirable to understand the convergence properties when summing over an infinite number of single trace operators. In order to do this, one would like to understand the weighted spectral density of single-trace operators, maybe along the lines of \cite{Pappadopulo:2012jk}. Subtleties may occur when summing over an infinite number of intermediate particles. It would be interesting to understand precisely which properties to require for the Mellin amplitude at order $1/N^2$. In addition to an infinite number of poles at finite $1/\sqrt{\lambda}$, one expects a specific behaviour for large Mellin variables, {\it e.g.} arising from the correct Regge behaviour. It may be simpler to propose a solution for the full Mellin amplitude at order $1/N^2$ (or even at finite $N$!), once all the conditions are specified, as opposed to obtain it as the sum of infinite contributions. 

The anomalous dimension of double trace operators $\gamma_{n,\ell}$ for large $n$ is related to certain singularities in the correlator, after analytic continuation to the Lorentzian regime. We have seen that for ${\cal N}=4$ SYM they take the form  (\ref{N4sing}). It would be very interesting to reproduce this singularity by alternative methods. For instance, it is known that the double null limit $u,v \to 0$ limit of this correlator is governed by the expectation value of a Wilson loop \cite{Alday:2010zy}. This result can also be derived from crossing symmetry \cite{Alday:2013cwa}. It would be interesting to make a similar statement for the bulk-point singularity.  

Finally, some of the ingredients of section three,{\it e.g.} the flat space limit of the Mellin amplitude leading to the Virasoro Shapiro amplitude, or the form of the OPE coefficient (\ref{OPElargetau}) for a large twist intermediate single-trace operator, required an input from the dual bulk theory. It would be interesting to recover these results from a purely CFT perspective.

\section*{Acknowledgments}

We are grateful to J. Penedones  for useful discussions. We thank the Galileo Galilei Institute for Theoretical Physics for the hospitality and the INFN for partial support during the completion of this work. The work of L.F.A was supported by ERC STG grant 306260. L.F.A. is a Wolfson Royal Society Research Merit Award holder. The work of A.B. is partially supported by Templeton Award 52476 of A. Strominger and by Simons Investigator Award from the Simons Foundation of X. Yin.

\appendix

\section{Mack polynomials}
\label{Mack}
The conformal blocks defined in the body of this paper depend on the Mack polynomials $P_\Delta^{(\ell)}(y,z)$. These are symmetric polynomials of total degree $\ell$ and were defined for instance in \cite{Mack:2009mi, Fitzpatrick:2011hu}. It is convenient to define them through a difference equation:

\begin{eqnarray}
\left(4(2+z)^2 t_y^- t_z^+ + 4 (y+2)^2 t_y^+ t_z^- + (\Delta+2x-\ell)(\Delta-2x+\ell+4)(t_y^- + t_z^-)\right) P_\Delta^{(\ell)}(y,z) = \nonumber \\
2(\Delta ^2+4 \Delta +\ell^2+2 \ell-2 y^2-8 y z-16 y-2 z^2-16 z-16)P_\Delta^{(\ell)}(y,z)
\end{eqnarray}
where we have defined the translation operators $t_y^\pm f(y) = f(y\pm1)$ and $t_z^\pm f(z) = f(z\pm1)$. This difference equation can be derived from the corresponding Casimir equation that the conformal block satisfies in space-time. This difference equation fixes fully the polynomial up to an overall normalisation factor. The normalisation factor can be fixed by requiring the correct small $u$ behaviour for the superconformal blocks. This leads to 

\begin{eqnarray}
P_{\Delta}^{(\ell)}(-2,\frac{\Delta-\ell}{2}) =-\frac{ 2^{-\ell-1} \csc (\pi  \Delta ) \Gamma (\ell+\Delta +4)}{\Gamma (-\Delta -2) \Gamma \left(\frac{1}{2} (-\ell+\Delta +4)\right)^2 \Gamma \left(\frac{1}{2} (\ell+\Delta +4)\right)^2}
\end{eqnarray}
Supplementing the above difference equation with this value fixes uniquely the polynomials. In this paper we will be interested in the "flat-space" limit of the Mack polynomials. In this limit we obtain:

\begin{equation}
P_{\ell+\tau}^{(\ell)}(\tau^2 y, \tau^2 z) = \frac{16 \times 4^{\tau+\ell} \tau^{4\ell}}{\pi^2} P_\ell(\theta)  \rho^\ell + \cdots
\end{equation}
where we have introduced $y= \rho(1-\cos \theta)$ and $z=\rho(1+\cos \theta)$. $P_\ell(\theta)$ are related to the Legendre polynomials and are given by
\begin{equation}
P_\ell(\theta) = \frac{\sin(\ell+1)\theta}{\sin \theta} 
\end{equation}
These functions are orthonormal with respect to the following measure:

\begin{equation}
\int_0^{2\pi} \frac{\sin^2\theta}{\pi} d\theta P_\ell(\theta)P_{\ell'}(\theta) = \delta_{\ell,\ell'}
\end{equation}

\section{Single poles vs unitarity}
\label{singlepoleunitarity}
In this appendix we briefly show that the presence of a single generic pole in the Mellin amplitude:

\begin{equation}
M_\delta(x) \sim \frac{1}{x+\delta/2}
\end{equation}
is not consistent with unitarity. Indeed, such a contribution corresponds to a primary scalar operator of dimension $\delta$, so that in space time:

\begin{equation}
A_\delta(u,v) = v^2 u^{\delta/2} g_{\delta+4,0}(u,v) + \cdots
\end{equation}
where we have assumed (by unitarity) that the corresponding OPE coefficient is positive, and we have normalised it to be one. However, descendants of that primary will lead to additional poles, which in order to be canceled require the exchange of scalar primaries with higher twist. In order to cancel all the poles except the first one, we need:

\begin{equation}
\label{singlepoleexpansion}
A_\delta(u,v) = v^2 \sum_{n=0}^\infty \alpha_n u^{\delta/2+n} g_{\delta+4+2n,0}(u,v)
\end{equation}
where

\begin{equation}
\alpha_n = \frac{\sqrt{\pi } (-1)^n \Gamma \left(\frac{\delta +5}{2}\right) 2^{-\delta -4 n-2} (\delta +2 n+2) \Gamma \left(n+\frac{\delta }{2}+2\right)^2 \Gamma (n+\delta +2)}{\Gamma \left(\frac{\delta }{2}+2\right)^3 \Gamma (n+1) \Gamma \left(n+\frac{\delta }{2}+\frac{3}{2}\right) \Gamma \left(n+\frac{\delta }{2}+\frac{5}{2}\right)}
\end{equation}
for generic, not even,  $\delta$ these are new operators and hence they should appear with a positive OPE coefficient, but we see that half the coefficients are actually negative! so a single pole is not consistent with unitarity. One may try to overcome this with a finite number of poles. However, we have found that for any finite number of poles the positivity condition for all OPE coefficients is too constraining. Hence we conclude that a finite number of generic poles is not consistent with unitarity. A similar analysis can be carried out for non scalar operators, but the analog of (\ref{singlepoleexpansion}) is much more complicated.

\end{document}